# Fokker-Planck equation for coupled anharmonic oscillators: symplectic integration and application to mean-field model of cooperative behavior

E. Klotins

*Institute of Solid State Physics, 8 Kengaraga Str., LV 1063 Riga, Latvia*

Abstract. Kinetics of systems specified by anharmonic, nonconservative and nonlocally coupled model Hamiltonians is reproduced in the Fokker-Planck and imaginary time Schrödinger equation techniques with subsequent symplectic integration. Test solutions focused on dielectric response in ferroelectrics demonstrates potential of this technique even for nonlocal and hardly nonlinear problems of polarization switching, and reassure the H-theorem of global stability [M. Shiino, Phys. Rev. A, Vol 36, pp. 2393-2411 (1987)] in the global coupling and zero driving limit.

1. Introduction

Canonical analysis of metastable systems is based on model Hamiltonians comprising both anharmonic and external driving terms [1,2]. Inspiring results are achieved as given by path-integral solutions [3 and references there], Floquet function approach [4], and the analytical eigenfunction technique [5]. The corresponding experimental situation is addressed to systems under alternate driving and additive noise: stochastic resonance [6,4],and noise activated sensors [7], to name only a few.

Nevertheless, conceptual and technical problems appear in going from the aforementioned local description comprising uniquely determined stationary states to the nonlocal description addressed to systems exhibiting ground-state bifurcation and divergence of relaxation time. Despite its importance for evolution of ferromagnetic /ferroelectric ordering, the relevant theory is still a challenge and involves an extension of canonical quartic and Ginzburg-Landau model Hamiltonians toward coupled nonlinear oscillator model [8]. The associated problems concern integration of nonlinear Fokker-Planck equation [9,10]and stability analysis of the solutions. The subject of present paper is the Fokker-Planck - imaginary time Schrödinger equation technique [5] combined with symplectic integration [10,11,12] attracting renewed attention in the context of large variety of physical contexts[8,11 ].This technique, inheriting most advantages from its quantum counterparts - norm conservation, long time stability, and the obtaining of the auxiliary function from which the observables can be computed - is essentially nonperturbative and capable for real time real space solutions. Early results for metastable systems modeling, in some extent, dynamic hysteresis are given in [13, 15, and 14]. The analytical part of computations is based on imaginary time Schrödinger equation derived from relevant Fokker-Planck equation making use of Wentzel-Kramer-Brilluin (WKB) type ansatz [5]. The numerical part is based on symplectic integration technique that transforms the problem to the real time recurrence relations. However, taking in game the cooperative behavior a more complex than the routine Ginzburg-Landau energy functional is necessary modeling a set of coupled anaharmonic oscillators [8]. An essential feature of relevant Fokker-Planck and imaginary time Schrödinger equations is strong nonlinearity which, nevertheless, is managed in nonperturbative fashion in the course of symplectic integration [9]. Whereas there is no first principle representation of this model [15], it returns most properties of interest and, above all, demonstrates capability of symplectic integration even for strongly nonlinear and nonlocal problems. The general framework of this kind of analysis is given in Sect.2 with special emphasis on the mapping between Fokker-Planck and imaginary time Schrödinger equation. Comparison between well accepted eigenfunction approach [4] and the symplectic integration of linear Fokker-Plank equation is illustrated in Sect.3. Symplectic integration of nonlinear Fokker-Planck equation is given in Sect.4 exhibiting fine details of polarization switching within the model of globally coupled anharmonic oscillators. Spatial extension within the framework of locally coupled anharmonic oscillators is worked out in Sect.5 with the central results Eq. (22) recovering the size effect on remnant polarization and symplectic integrator Eq. (25) for instantaneous polarization. Features and possible extension of this technique is discussed in Sect.5. Final conclusions are presented in Sect.6.

## 2. Essentials to symplectic integration of imaginary-time Schrödinger equation

This section starts with an outline of the Fokker-Planck and imaginary time Schrödinger equation mapping for the standard Ginzburg-Landau model Hamiltonian [1].Though this technique is not new [5,16]a briefly review is motivated to make the further analysis self-contained. The dimensionless Fokker-Planck equation for probability density of polarization $\rho(P,t)$ reads as

$$\frac{\partial \rho(P,t)}{\partial t} = \frac{\partial}{\partial P}\left(\frac{\delta U}{\delta P}\rho(P,t)\right) + \Theta\frac{\partial^2 \rho(P,t)}{\partial P^2} \tag{1}$$

Here

$$U = -P^2/2 + P^4/4 + (\nabla P)^2/2 - \lambda(t)P, \tag{2}$$

is the energy functional, the diffusion coefficient $\Theta$ bounds together parameters of the system, $\lambda(t)$ specify the time dependent external field, and the gradient term $(\nabla P)^2$ specify weak nonlocality. Also the additive zero mean Gaussian thermal noise is assumed. The multivariate nature of the probability density $\rho(\{P\},t)$ is not exposed explicitly as motivated by its later splitting in statistically independent parts over the spatial mesh. The concept is to transform Eqs.(1,2) in imaginary time Schrödinger equation for subsequent symplectic integration. Omitting temporary the gradient term in Eq.(2) yields quartic potential for which the standard WKB ansatz $\rho(P,t) = \exp[F(P)]G(P,t)$ [5] produces imaginary time Schrödinger equation

$$\frac{\partial G(P,t)}{\partial t} = \left[\Theta \frac{\partial^2}{\partial P^2} + V(P)\right]G(P,t) \tag{3}$$

In Eq.(3) the potential operator $V(P)$ reads as

$$V(P) = \left[-\frac{1}{4\Theta}[U'(P)]^2 + \frac{1}{2}U''(P)\right] \tag{4}$$

and the auxiliary function $G(P,t)$ given by Eq.(3) unfolds polarization kinetics through the first moment of probability density $\rho(P,t)$. The key part of computations includes analytical solution of an ordinary differential equation for $F(P)$ canceling the first derivative of auxiliary function in Eq.(3) and simultaneously determining the WKB ansatz as

$$\rho(P,t) = \exp[-U(P)/2\Theta]G(P,t) \tag{5}$$

The mapping of Eqs.(1,3) is quite general and applicable also for nonlocal energy functionals. The analytical and quite exact part of computations is completed by recurrence relation for the auxiliary function valid for a small time slice $\Delta t$

$$G(P, t+\Delta t) = \exp\left[\Delta t\left(\Theta \frac{\partial^2}{\partial P^2} + V(P)\right)\right]G(P,t) \tag{6}$$

The symplectic integrator for Eq.(5) derived in **Appendix A** reads as

$$\left(1 - \frac{\Theta \Delta t}{2}\frac{\partial^2}{\partial P^2}\right)G(P, t+\Delta t)$$
$$= \left\{\exp\left[\frac{\Delta t}{2}V + \frac{\Delta t^3}{48}(\nabla V)^2\right]\left(1 + \frac{\Theta \Delta t}{2}\frac{\partial^2}{\partial P^2}\right)\exp\left[\frac{\Delta t}{2}V + \frac{\Delta t^3}{48}(\nabla V)^2\right]\right\}G(P,t) \tag{7}$$

Here the potential operator $V$ is given by Eq.(A5) with time argument $t := t + \frac{\Delta t}{2}$. Advantages of the symplectic integrator Eq.(7) are norm conservation, perfect stability, and suitability for non-conservative energy functionals Eq.(2).

3. Dynamic hysteresis

Dynamic hysteresis appears in metastable systems as a combined effect of periodic driving and thermal noise. In the absence of external driving the metastable states are symmetric and the observable exhibit a unique ground state. Otherwise, if the rate of external driving is lesser than another intrinsic time-scale, the equilibrium holds at the instantaneous value of the driving. In general, this adiabatic approximation does not hold since the population of metastable states cannot follow the external forcing and, in the case of periodic driving, exhibit itself as the dynamic hysteresis. A prototypical setup for this kind of analysis is an overdamped Brownian particle in presence of periodically driving and thermal noise. Considering thermal noise as the Wiener process, the system is Markovian and time evolution of probability distribution is given by Fokker-Planck equation for a particular potential. If the time of integration substantially exceeds the period of external driving, a very accurate stationary solution for quartic potential is available [4].Details of this technique with application to dynamic hysteresis are given in [13].The results in Fig.1 confirm, for a representative set of parameter values, that the agreement between the solutions[4], [13] and the test solution results preceded by Eq.(5) for the dynamic hysteresis is fairly good and reassures accuracy of the rather complex approach Eqs.(1-6).In more detail the solution [4] is semiadiabatic in the sense that only the first non-zero eigenvalue contributes in the dynamics. Otherwise, no explicit time scales are defined for the Fokker-Planck imaginary time Schrödinger approach Eqs. (1-6). In this context the change of parameters initiated by harmonic external field $\lambda(t) \propto \sin(\Omega t)$ at $\Omega = 10^{-3}$ dimensionless driving frequency is slow compared to other intrinsic relaxation times.

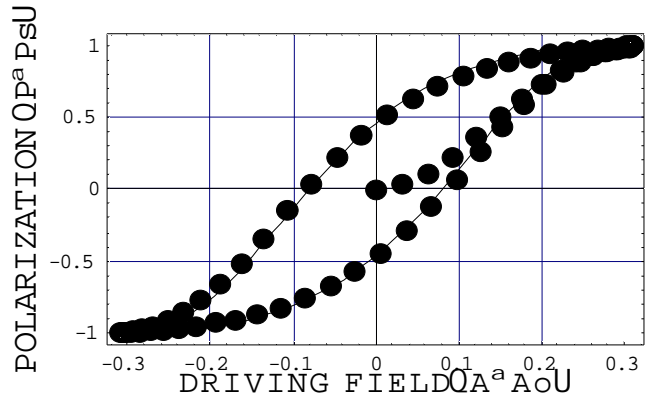

Fig.1  Dynamic hysteresis under harmonic field: semiadiabatic approach [2] (line) and symplectic integration (dots). Parameters of the problem: amplitude of the external field $\lambda = 0.309\lambda_0$, frequency $\Omega = 10^{-3}$, $\Theta - 1/20$, $\lambda(t) = \sin(\Omega t)$, and $\lambda_0 = 2/\sqrt{27}$ is the static coercive field.

## 4. Polarization switching in the model of globally coupled anharmonic oscillators

Standard mean field approach that is modeled by Ginzburg-Landau energy functional Eq. (2) is not rich enough to generate effects associated with bifurcation of the ground state and divergence of the relaxation time. For this study the globally coupled model is introduced [8] as defined in terms of $N-$particle Markov system for a set of overdamped anharmonic oscillators interacting by way of attractive linear coupling and affected by additive thermal noise. In this globally coupled model the interaction does not depend on the distance between elements. On physical grounds this extension of Ginzburg-Landau energy functional Eq.(2) concern a set of $N$ interacting subsystems, each of them characterized by a single degree of freedom and whose dynamics is governed by Langevin equations

$$\frac{\partial P_i}{\partial t} = -\frac{\partial F(P_i)}{\partial P_i} + \sum_{k=1}^{N}\frac{\varepsilon}{N}(P_k - P_i) + \eta_i(t) \tag{8}$$

Here $\langle \eta_i(t)\eta_j(t')\rangle = \delta_{ij}\delta(t-t')$ specify the statistically independent white noise, and the constant $\varepsilon > 0$ denotes the strength of attractive mean-field type coupling (at $N=1$ Eq.(8) reduces to Eq.(2)). At the thermodynamic $N \to \infty$ limit the averages of $P_k$ in Eq. (8) can be assumed to behave in a deterministic way, namely, $\lim_{N\to\infty}\left(\frac{1}{N}\sum_{k=1}^{N}P_k(t)\right) = \overline{P}(t)$ and the corresponding Fokker-Planck equation concern probability density for each $P_i$ which originates from various realizations of white noise

$$\frac{\partial \rho}{\partial t} = \sum_{i=1}^{N}\left[-\frac{\partial}{\partial P_i}\left[-\frac{\partial F}{\partial P_i} + \frac{\varepsilon}{N}\sum_{k=1}^{N}P_k - \frac{\varepsilon}{N}\sum_{k=1}^{N}P_i\right]\rho + \Theta\frac{\partial^2 \rho}{\partial P_i^2}\right] \tag{9}$$

Recognizing that $\frac{\varepsilon}{N}\sum_{k=1}^{N}P_k(t) = \varepsilon\overline{P}(t)$ and each $i-$th block is described by equal kinetics Eq.(8) the Eq.(9) reduces to

$$\dot{\rho}(P,t) = \frac{\partial}{\partial P}\left[U'(P,t) + \Theta\frac{\partial}{\partial P}\right]\rho(P,t) \tag{10}$$

For a potential $U(P,t) = -\frac{P^2}{2} + \frac{P^4}{4} + \frac{\varepsilon}{2}[P - \overline{P}(t)]^2 - P\lambda(t)$ it is convenient to assign the local terms by $U_1(P,t) = -\frac{P^2}{2} + \frac{P^4}{4} - P\lambda(t)$ and to maintain the nonlinear term in its explicit form $U_2(P,t) = \frac{\varepsilon}{2}[P - \overline{P}(t)]^2$. It yields the stationary solution (assumed as the initial condition for further calculations) of Eq.(10) as

$$\rho(P,0) = \frac{\exp\left[-\dfrac{U_1(P,0)+U_2(P,0)}{\Theta}\right]}{\int \exp\left[-\dfrac{U_1(P,0)+U_2(P,0)}{\Theta}\right]dP} \qquad (11)$$

Here the denominator provides normalization of the probability distribution $\int \rho(P,0)dP = 1$ and the stationary (initial) value of the first moment of polarization density $\overline{P}(0)$ is found by integrating Eq.(10) over $P$ with $\overline{P}(0)$ as a parameter. In $P, P - \overline{P}(0)$ frame the exact value of $\overline{P}(0)$ is found as intersection of $P - \overline{P}(0)$ plot with the $P$ axis as shown in Fig.2

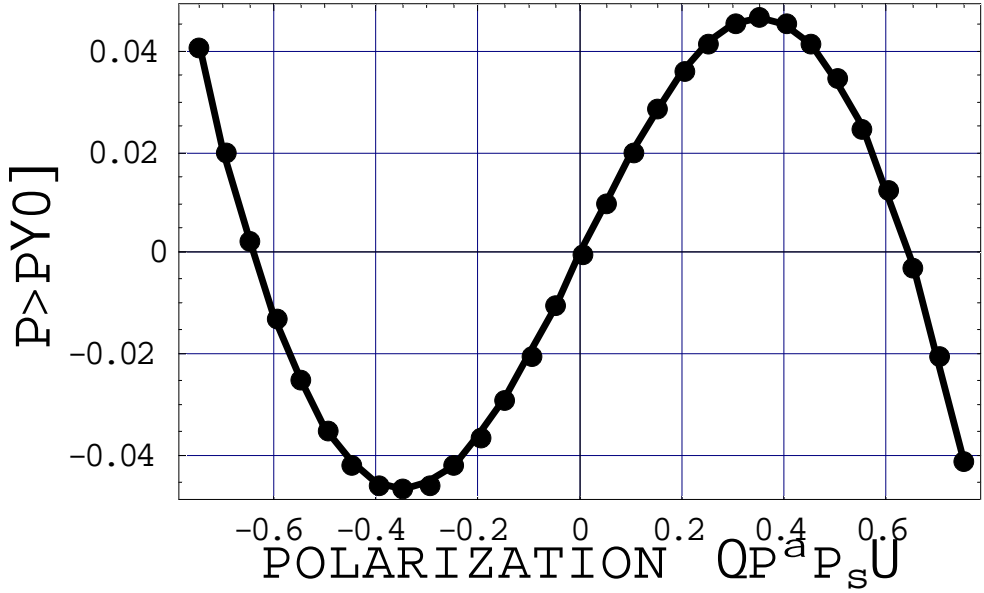

Fig. 2 Schematic plot of $P - \overline{P}(0)$ as a function of polarization $P$ (in $P_s$ units). The stationary polarization $\overline{P}(0)$ is recovered by intersection of $P - \overline{P}(0)$ plot with the $P$ axis exhibiting two stable solutions at $\overline{P} = \pm 0.64$ and an unstable solution at $\overline{P} = 0$. Parameters of the model: $\varepsilon = 0.07$, $\lambda = 0$, $\Theta = 1/20$.

Objective is the nonstationary solution of nonlinear Fokker-Planck equation Eq. (10). Unlike solutions for linear Fokker-Planck equations based on the Boltzmann's H-theorem (which ensures the existence of a uniquely determined long-time probability distribution $\rho_\infty(P,t)$) and the Floquet theorem (which provides

$\rho_\infty(P,t)$ the same period as the external driving) [4] there are no analogues of these theorems in case of nonlinearity. Exception is [8] stating that (at overcritical interaction constant) the system always reaches global stability in the sense that there is no other attractors corresponding to the stationary solution Eq.(10) and that any time dependent solutions of Eq.(10) lying far from equilibrium must be attracted by either one of those stationary solutions without any possibility of runaway behavior or limit cycle type oscillations. A linear response theory is derived in [17]. The Fokker-Planck – imaginary time Schrödinger scheme preceded hereafter recovers all features of global stability [8] as well as the bifurcation between attractors (specified by two remnant polarizations). Modeling of temporal behavior of the probability density Eq.(10) starts with the ansatz derived in Appendix B

$$\rho(P,t) = \exp\left[-\frac{U_1(P,t)+U_2(P,t)}{2\Theta}\right]G(P,t) \qquad (12)$$

and the auxiliary function $G(P,t)$ given by nonlinear imaginary time Schrödinger equation

$$\dot{G}(P,t) = \left[\Theta\frac{\partial^2}{\partial P^2} + V_1(P,t) + V_2(\overline{P}(t),P,t)\right]G(P,t) \qquad (13)$$

made up of both the linear $V_1(P,t)$ and the nonlinear $V_2(\overline{P}(t),P,t)$ terms in the potential operator. Here

$$V_1(P,t) = -\frac{1}{4\Theta}\left[\frac{\partial U_1(P,t)}{\partial P}\right]^2 + \frac{1}{2}\frac{\partial^2 U_1(P,t)}{\partial P^2} + \frac{1}{2\Theta}\left[\frac{\partial U_1(P,t)}{\partial t}\right] \qquad (14)$$

and

$$V_2(\overline{P}(t),P,t) = \frac{1}{2\Theta}\frac{\partial U_2(P,t)}{\partial t} - \frac{1}{2\Theta}\frac{\partial U_1(P,t)}{\partial P}\frac{\partial U_2(P,t)}{\partial P} - \frac{1}{4\Theta}\left(\frac{\partial U_2(P,t)}{\partial P}\right)^2 + \frac{1}{2}\frac{\partial^2 U_2(P,t)}{\partial P^2} \qquad (15)$$

Series expansion of the first moment is given by

$$\overline{P}(t) = \overline{P}(t_0) + Q(t-t_0) + \ldots \qquad (16)$$

here $t_0 = (n-1)\Delta t$, $n$ is the number of recursion, $\Delta t$ is the time increment, and the dynamic nature of potential is accounted for by setting $t := t + \Delta t/2$ [10]. Computing details concern polarization mesh $P_i = P_{\min} + (i-1)\Delta P$, $i \in [1, N+1]$ (here the polarization increment $\Delta P = (P_{\max} - P_{\min})/N$), and transforming the exponential operator [18] Eq.(A6) in two 10-diagonal matrices. Another essential detail is the merit $M(Q) = \int \rho(P,Q,t)dP - (\overline{P}(0) + Q\Delta t)$ found as analytical function of $Q$ by interpolation. The polarization rate $q$ fitting $M(q)=0$ is put into operation in the final (fourth) step when Eq. (13) is re-evaluated at $Q:=q$ and the final auxiliary function $G(\{P\},t+\Delta t)$ found. The recurrence step completes with relation $\overline{P}(t_i+\Delta t):=\overline{P}(t_i)+q_i\Delta t$. Since no analytical estimates are known for these calculations error control and accuracy tests are preceded for temporal deviation of the stationary states, normalization of the probability distribution, polarization rate, and the norm conservation. Results illustrated in Appendix C reassure the global stability found in [8]. Assurance of the accuracy and stability of this approach Eqs.(10-16) makes available to model the polarization switching, namely, a monopolar pulse controlled transition from one stationary state to another.

Figure 3 illustrates polarization response of a system with initially positive remnant polarization (as shown in Fig.5). The driving is preceded by a negative sow teeth shaped pulse specified by -0.1, - 0.05, and -0.027 dimensionless amplitudes scaled after $2/\sqrt{27}$ static coercive field as accepted in [4,13,15]. In more detail the $-0.1$ driving field corresponds $\sim 1/4$ of the thermodynamic coercive field. For each amplitude of the driving field two varieties of polarization responses are modeled for $100\pi$ (short pulse) and $1000(\pi/2)$ (moderate pulse), correspondingly. The dimensionless pulse lengths are harmonized with the frequency scale accepted in [4,13,15] and corresponds to $\Omega=1/100$ for short pulses $\Omega=1/500$ for moderate pulses. In these terms the sow teeth shaped pulse reduces linearly within $t\in[0,\pi/2\Omega]$, grows linearly within $t\in[\pi/2\Omega,\pi/\Omega]$, and holds at zero for $t>\pi/\Omega$.

What is well accepted for this experimental situation is the H-theorem of global stability [8] stating that there is not other attractor than set of the Eq. (10). It means that the sign of remnant polarization approached by the system depends on the sign of the first moment of probability distribution $\overline{P}(\pi/\Omega)$ at the time instant the driving is switched off, namely, $\overline{P}(\pi/\Omega)>0$ yields $\overline{P}(\infty)\to\overline{P}(0)$, and $\overline{P}(\infty)\to-\overline{P}(0)$ otherwise. The first case is confirmed by the $-0.025$ a.u. short pulse (dots). Otherwise, the polarization plot corresponding to $-0.025$ a.u. moderate length driving pulse (line) and exhibiting $\overline{P}(\pi/\Omega)<-\overline{P}(0)$ approaches to $-\overline{P}(0)$ from the bottom side in accord with the H-theorem of global stability. Similarly, the polarization plot corresponding to $-0.05$ a.u. short pulse (dots) exhibit $\overline{P}(\pi/\Omega)<0$ and approaches to $-\overline{P}(0)$ from the top side also in accord with this H-theorem. All amplitudes exceeding $0.05$ a.u. are obviously overcritical as manifested by the rest of plots illustrating the polarization switching. Whereas the solution returns the H-theorem of global stability at zero driving, its generalization for nonzero driving is an open problem. For this representative set of parameters the driving-polarization response is rather hysteretic.

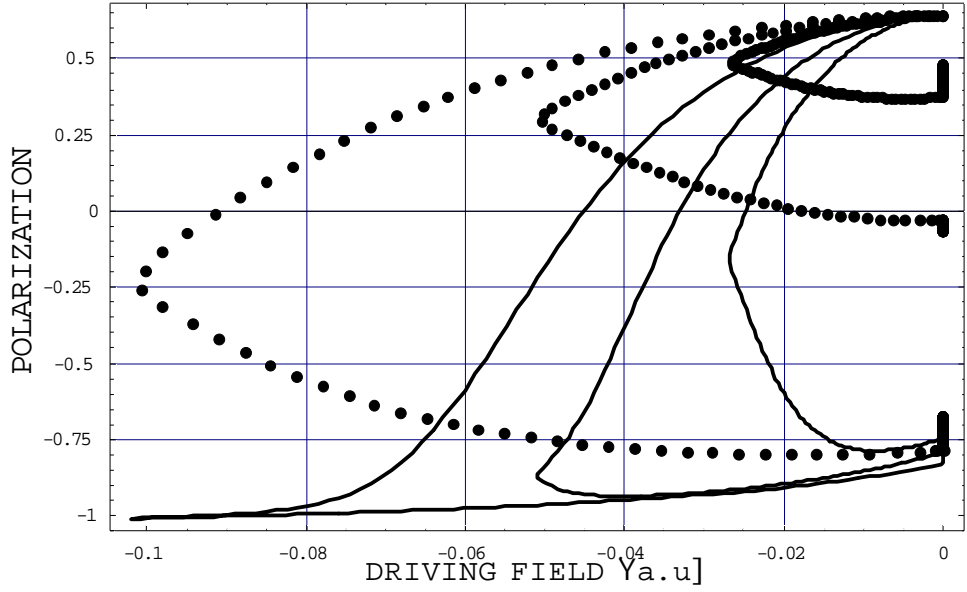

Fig. 3 Polarization switching from $\overline{P}_r(0) = +0.64$ stationary (initial) state to the $\overline{P}_r(0) = -0.64$ final state initiated by -0.1, - 0.05, and -0.027 sow teeth shaped pulses of short (dots) and moderate (line) length.

Aforementioned analysis of polarization response is paralleled with temporary behavior of the occupancy of polarization states $\Xi(t) = \int_{P(\rho_{min})}^{\infty} \rho(P,t) dP$ (here $P(\rho_{min})$ is the polarization, corresponding to the minima of the bimodal probability distribution). For the given set of parameters $\Xi \in [0.836, 1-0.836]$ and the upper bound corresponds to positive remnant polarization in Fig.2. Figure 4 illustrates time propagation of the occupancy. The vertical $157, 785$ and $314, 1570$ a.u. grid lines points the instants of maximum amplitude and the pulse length for short (bold dots) and moderate length pulses (lines), correspondingly. Under moderate length pulses the occupancy reduces down to $\Xi(t \to \infty) = 1 - 0.836$ in nonmonotonously fashion, namely, both the minimum occupancy and its $\Xi(t \to \infty) = 1 - 0.836$ stationary value is delayed which respect to the driving pulse and the system approaches to $-\overline{P}(0)$ (and $1-0.836$ occupancy) from the bottom side. This behavior is different in case of short pulses (bold dots). Firstly, the occupancy is substantially delayed and its minima appear only at the instant of pulse length. Secondly, after the driving switched off, the occupancy enlarges in case at minimum amplitude and reduces at the rest.

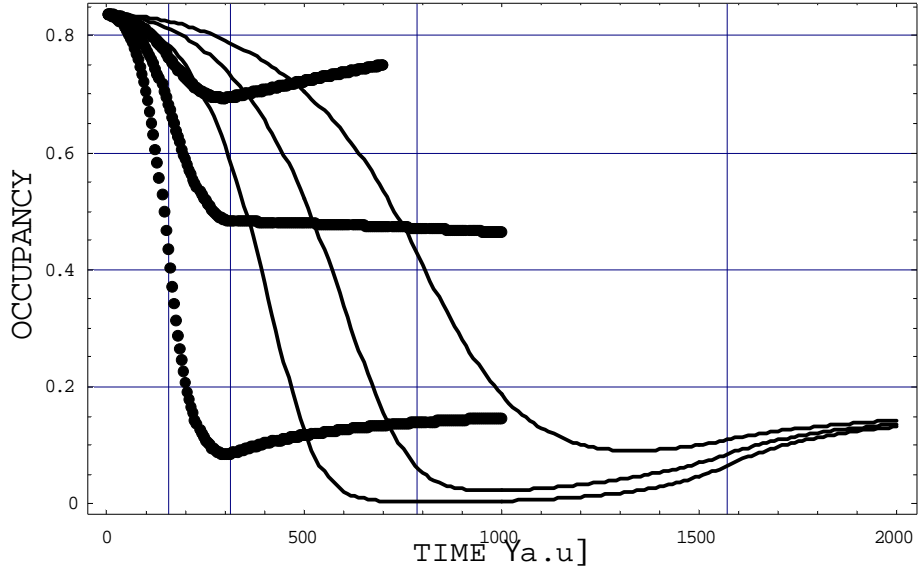

Fig.4 Time propagation of occupancy modeled for short sow teeth shaped pulses (bold dots) and moderate driving pulses (lines) specified by various amplitudes.

What is missing in this analysis is the spatial extension lost in the case of globally coupling. Nevertheless, the aforementioned mathematical technique, namely, the implementation of recursion-specific first moment $\overline{P}(t_i)$ Eq.(23) has essential consequences for spatially extended problem emerged by locally coupled model in which each anharmonic oscillator is coupled which its first neighbors.

5. Polarization switching in the model of locally coupled anharmonic oscillators

Spatial dependence of polarization field, disappearing in the model of globally coupled anharmonic oscillators Eqs.(8) may be restored ad hoc considering the model of locally coupled oscillators for which the interaction terms are restricted to first neighbors. This approach assumes that (i) the system consist of finite number of (microscopically large) blocks modeled by Ginzburg-Landay energy functional $\Phi_i = -\frac{1}{2}P_i^2 + \frac{1}{4}P_i^4 - \lambda(t)P_i$ and (ii) the first neighbor interaction between (macroscopically small) blocks holds so addressing the problem to ensemble of interacting blocks. Going around the microscopic interpretation of the strength of interaction and the correlation length, the problem is preceded by introducing a positive coupling acting at the distance of spatial increment. In this case the energy is given by the model Hamiltonian

$$H \equiv \sum_i^N \left\{ \Phi_i + \frac{\varepsilon}{2}\left( \left(\overline{P}_{i+1}(t) - P_i\right)^2 + \left(\overline{P}_{i-1}(t) - P_i\right)^2 \right) \right\} \qquad (17)$$

Here $\Phi_i = -\frac{1}{2}P_i^2 + \frac{1}{4}P_i^4 - \lambda(t)P_i$, the first moments of probability density $\overline{P}_k$ are unknown quantities and are evaluated selfconsistently afterward. Kinetic equations derived from Eq.(17)

$$\frac{\partial P_i}{\partial t} = -\frac{\partial \Phi_i}{\partial P_i} + \varepsilon\left(\overline{P}_{i+1}(t) - 2P_i + \overline{P}_{i-1}(t)\right) \qquad (18)$$

readdress the problem to a set of Fokker-Planck equations

$$\dot{\rho}(P_i,t) = -\frac{\partial}{\partial P_i}\left[-\frac{\partial \Phi_i}{\partial P_i}\rho(P_i,t) + \varepsilon\left(\overline{P}_{i+1}(t) - 2P_i + \overline{P}_{i-1}(t)\right)\rho(P_i,t)\right] + \Theta_i \frac{\partial^2}{\partial P_i^2}\rho(P_i,t) \qquad (19)$$

In stationary case

$$\rho(P_i)\left(2\varepsilon + \frac{\partial \Phi_i}{\partial P_i^2}\right) + \left(\varepsilon\left(\overline{P}_{i-1} - 2P_i + \overline{P}_{i+1}\right) + \frac{\partial \Phi_i}{\partial P_i}\right)\frac{\partial \rho(P_i)}{\partial P_i} + \Theta\frac{\partial^2 \rho(P_i)}{\partial P_i^2} = 0 \qquad (20)$$

and the stationary probability density yields

$$\rho(P_i) = C\exp\left[\frac{-\Phi(P_i) + \varepsilon P_i(\overline{P}_{i-1} - P_i + \overline{P}_{i+1})}{\Theta_i}\right] \qquad (21)$$

Here $C$ is normalization constant, and $\overline{P}_0 = 0$, $\overline{P}_{i_{max}+1} = 0$ are zero boundary conditions. Implementing normalization of the probability density as well as the first moment $\overline{P} = \int_{-\infty}^{\infty} P\rho dP$ the selfconsistency condition for $\overline{P}_i$ is given by

$$\frac{\int_{P_{min}}^{P_{max}} P_i \exp\left[\frac{-\Phi(P_i,0) + \varepsilon P_i(\overline{P}_{i-1} - P_i + \overline{P}_{i+1})}{\Theta_i}\right]dP_i}{\int_{P_{min}}^{P_{max}} \exp\left[\frac{-\Phi(P_i,0) + \varepsilon P_i(\overline{P}_{i-1} - P_i + \overline{P}_{i+1})}{\Theta_i}\right]dP_i} - \overline{P}_i = 0 \qquad (22)$$

For a set of starting values obtained, for example, from a static approach [19] Eq. (22) gives the stationary solution of Cauchy problem. Figure 10 demonstrates example solutions of Eq.(22) for zero boundary conditions, $\overline{P} = \pm 1$ starting values, and various coupling between blocks as a parameter.

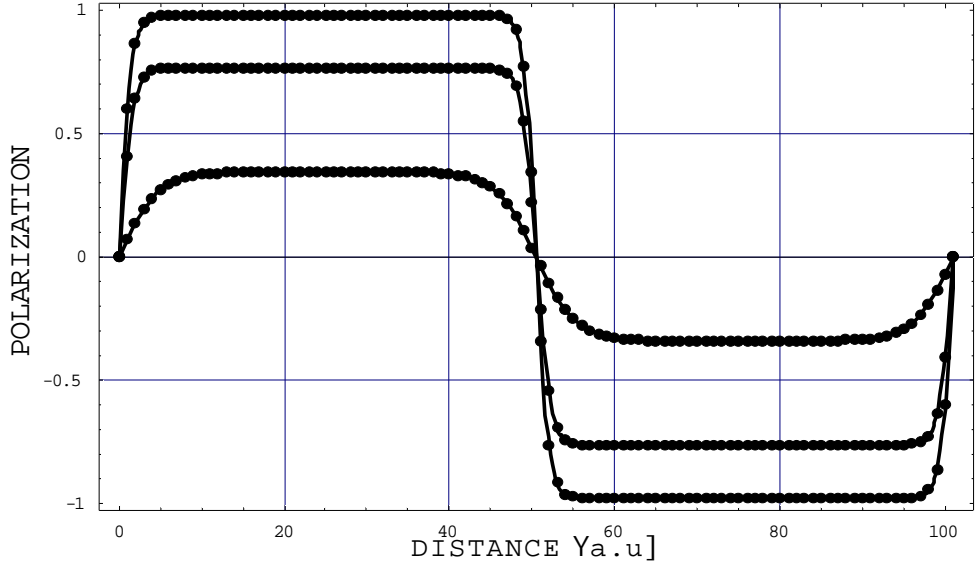

Fig. 5 Example solution for $180^0$ domains in a 1-D region with zero boundary conditions and $0.3, 0.4, 1$ coupling constants. For $\varepsilon \to 1$ coupling constant the difference between the spontaneous and remnant polarization disappears as illustrated by the upper plot.

Nonstationary solution of Eq.(20) is similarly to this illustrated in Appendix B and starts with the ansatz

$$\rho(P_i, t) = \exp[-F(P_i, t)] G(P_i, t) \tag{23}$$

Inserting the ansatz Eq. (23) in Fokker-Planck equations Eqs. (19) and zeroing the coefficients at $\partial G(P_i, t)/\partial P_i$ yields the relation for exponential factor in Eq.(23)

$$F_i = \frac{\varepsilon P_i (\overline{P}_{i-1} - P_i + \overline{P}_{i+1}) - \Phi(P_i)}{2\Theta} \tag{24}$$

Substitution of Eq.(24) in Eqs.(23,19) gives the imaginary time Schrödinger equation for the auxiliary function

$$\dot{G}(P_i, t) = [T[i] + V_1[i] + V_2[i] + K[i]] G(P_i, t) \tag{25}$$

Over a spatial mesh $i \in [1, i_{max}]$ the kinetic operator $T[i]$ is given by

$$T[i] = \Theta \frac{\partial^2}{\partial P_i^2} \tag{26}$$

Here the linear part $V_1[i]$ and the nonlinear part $V_2[i]$ of potential operators in Eq.(25) are given by

$$V_1[i] = -\frac{1}{4\Theta_i}\left(\frac{\partial \Phi_i}{\partial P_i}\right)^2 + \frac{1}{2}\left(\frac{\partial^2 \Phi_i}{\partial P_i^2}\right) \tag{27}$$

$$V_2[i] = \frac{\varepsilon\left(4\Theta_i - (2P_i - \overline{P}_{i-1}(t) - \overline{P}_{i+1}(t))\left(2P_i - \varepsilon(\overline{P}_{i-1}(t) + \overline{P}_{i+1}(t)) + 2\frac{\partial \Phi}{\partial P_i}\right)\right)}{4\Theta_i} \quad (28)$$

and the correction to the potential operators generated by explicit time dependence of the energy functional yields as

$$K[i] = \frac{-\varepsilon P_i \dot{\overline{P}}_{i-1}(t) - \varepsilon P_i \dot{\overline{P}}_{i+1}(t) + \dot{\Phi}_i}{2\Theta} \quad (29)$$

Aforementioned analytical calculations are followed by the numerical part comprising the solution of Eq.(25) and evaluation of the merit $M(Q) = \int P\rho(P,Q,t)dP - (\overline{P}(0) + Q\Delta t)$ (transformed in analytical function of $Q_i$ by quadratic interpolation). This trick generates a set of coupled algebraic equations $M(Q_i) = 0$ for expansion coefficients $Q_i$ so returning the density distributions by Eq.(23) over spatial mesh in every time slice. The relevant large scale computer modeling is out of scope of this work.

6. Discussion

Thermal noise activated nonadiabatic behavior of metastable systems is investigated in the context of electric hysteresis and polarization switching in ferroelectrics. Main focus is made on the mathematical technique as based on the Langevin – Fokker-Planck – imaginary time Schrödinger scheme addressing the problem to matrix recurrence relations in the course of symplectic integration. The definition of physical system is given by energy functionals of growing complexity and the test solutions captures finite size, spatial inhomogeneity, bifurcation of stationary states, and arbitrary alternate driving.
An instructive example is based on quartic energy functional and exposes features of the fundamental correctable second order propagator providing long term stability of the symplectic integration. Dynamic hysteresis modeled within this approach for periodic driving coincides with well accepted recent results found in eigenfunction technique.
Bifurcation of stationary states and divergence of relaxation time are captured by a more realistic energy functional for an assembly of coarse grained particles with attractive first neighbor interaction. This strictly nonlinear case is handled by series expansion of the first moments of probability density and readdresses evaluation of its coefficients to an algebraic problem. Polarization switching is modeled within this approach.

This trick allows spatial extension and makes the density distribution in different spatial points statistically independent. Based on this property the solution of complete Cauchy problem is demonstrated by a third model energy functional with interaction terms restricted to first neighbors so maintaining the spatial inhomogeneity of polarization field.

Concluding, it is shown how the Langevin, Fokker-Planck and imaginary time Schrödinger equation techniques can be derived elegantly in terms of symplectic integration even for nonlocal and hardly nonlinear problems and can be used in calculations of response properties of spatially inhomogeneous metastable systems.

## Acknowledgements

This work has been supported by the Contract No. ICA1-CT-2000-70007 of European Excellence Center of Advanced Material Research and Technology (Riga)

## Appendix A

Equation 5 can be considered as a formal solution for a separable and autonomous Hamiltonian system, for which the quantities $\Theta \frac{\partial^2}{\partial P^2}$ and $V(P)$ are non-commutative differential operators

$$\exp\left[\Delta t\left(\Theta \frac{\partial^2}{\partial P^2} + V(P)\right)\right] \neq \exp\left[\Delta t \Theta \frac{\partial^2}{\partial P^2}\right]\exp[\Delta t V(P)] \tag{A1}$$

Correct expression for Eq.(A1) is given by [10,12] the decomposition relation

$$\exp\left[\Delta t\left(\Theta \frac{\partial^2}{\partial P^2} + V(P)\right)\right] \approx \prod_{i=1}^{k}\exp\left[\Delta t c_i \Theta \frac{\partial^2}{\partial P^2}\right]\exp[\Delta t d_i V(P)] \tag{A2}$$

The symplectic integrator is then

$$G(P,\Delta t) = \left[\prod_{i=1}^{k}\exp\left[\Delta t c_i \Theta \frac{\partial^2}{\partial P^2}\right]\exp[\Delta t d_i V(P)]\right]G(P,t=0) \tag{A3}$$

where the operators are applied in the order of the factorization coefficients $(c_1, d_1, ..., c_k, d_k)$. For second order symplectic integrator adopted in this work $k=2$, $c_1=0$, $c_2=1$, $d_1=d_2=\frac{1}{2}$, and the Eq.(A3) yields

$$G(P, t+\Delta t) = \left\{ \exp\left[\frac{\Delta t}{2} V(P)\right] \exp\left[\Delta t \Theta \frac{\partial^2}{\partial P^2}\right] \exp\left[\frac{\Delta t}{2} V(P)\right] \right\} G(P,t) \tag{A4}$$

What is missing in Eq.(4) is the explicit time dependence of the potential operator $V(P)$, initiation of a symplectic corrector algorithm, and approximation for the operator $\exp\left[\Delta t \Theta \frac{\partial^2}{\partial P^2}\right]$. The time dependence is implemented in two steps. Firstly, the potential operator Eq.(4) is extended by a supplementary $\tilde{\lambda}(t)$ term, generated in course of WKB mapping

$$V(P,t) = \left[ -\frac{1}{4\Theta}[U'(P,t)]^2 + \frac{1}{2} U''(P,t) - \frac{P\tilde{\lambda}(t)}{2\Theta} \right] \tag{A5}$$

Secondly, the time argument in the second order decomposition Eq.(A2) is introduced as $V(P) := V\left(P, t + \frac{\Delta t}{2}\right)$. The symplectic corrector algorithm is given by $\frac{\Delta t^3}{48}(\nabla V)^2$ term supplementary to $\frac{\Delta t}{2} V(P)$ one in Eq.(A4). [11]

The operator $\exp\left[\Delta t \Theta \frac{\partial^2}{\partial P^2}\right]$ is approximated by its Cayley's form [20]

$$\exp\left[\Delta t \Theta \frac{\partial^2}{\partial P^2}\right] \cong \frac{1 + \frac{\Theta \Delta t}{2} \frac{\partial^2}{\partial P^2}}{1 - \frac{\Theta \Delta t}{2} \frac{\partial^2}{\partial P^2}} \tag{A6}$$

Finally, substituting Eq.(A6) for Eq.(A4) and moving the denominator onto the left-hand side yields

$$\left(1 - \frac{\Theta \Delta t}{2} \frac{\partial^2}{\partial P^2}\right) G(P, t + \Delta t)$$
$$= \left\{ \exp\left[\frac{\Delta t}{2} V + \frac{\Delta t^3}{48}(\nabla V)^2\right] \left(1 + \frac{\Theta \Delta t}{2} \frac{\partial^2}{\partial P^2}\right) \exp\left[\frac{\Delta t}{2} V + \frac{\Delta t^3}{48}(\nabla V)^2\right] \right\} G(P,t) \tag{A7}$$

Here the potential operator is given by Eq.(A5) with time argument $t := t + \frac{\Delta t}{2}$. Eq.(A7) is applicable for non-conservative energy functionals Eq.(2), display norm conservation property, and provides only oscillatory and not secular errors in the course of integration.

Appendix B

In terms of $U_1(P,t) = -\frac{P^2}{2} + \frac{P^4}{4} - P\lambda(t)$, $U_2(P,t) = \frac{\varepsilon}{2}[P - \overline{P}(t)]^2$ the Fokker-Planck equation Eq.(17) yields

$$\dot{\rho}(P,t) = (U_1'(P,t) + U_2'(P,t))\rho'(P,t) + \rho(P,t)(U_1''(P,t) + U_2''(P,t)) + \Theta\rho''(P,t) \quad (B1)$$

The ansatz mapping Eq.(B1) with imaginary time Schrödinger equation comprises the auxiliary function $G(P,t)$ and the exponential factor $F(P,t)$

$$\rho(P,t) = \exp[-F(P,t)]G(P,t) \quad (B2)$$

Substitution Eq.(B2) in Eq.(B1) yields a second order partial differential equation for the auxiliary function $G(P,t)$. With imaginary time Schrödinger equation in mind the relation for the exponential factor is found by condition, canceling first order derivative of $G(P,t)$, or explicitly,

$$-2\Theta F'(P,t) + U_1'(P,t) + U_2'(P,t) = 0 \quad (B3)$$

Hence

$$F(P,t) = \frac{U_1(P,t) + U_2(P,t)}{2\Theta} \quad (B4)$$

is the exponential factor and the WKB-type ansatz is given by

$$\rho(P,t) = \exp\left[-\frac{U_1(P,t) + U_2(P,t)}{2\Theta}\right]G(P,t) \quad (B5)$$

Inserting Eq.(B5) in Eq.(B1) gives imaginary time Schrödinger equation for the auxiliary function

$$\dot{G}(P,t) = \Theta G''(P,t) + G(P,t)\left[\begin{array}{c}\dfrac{\dot{U}_1(P,t) + \dot{U}_1(P,t)}{2\Theta} - \dfrac{[U_1'(P,t)]^2}{4\Theta} - \dfrac{U_1'(P,t)U_2'(P,t)}{2\Theta} \\ -\dfrac{[U_2'(P,t)]^2}{4\Theta} + \dfrac{U_1''(P,t) + U_2''(P,t)}{2}\end{array}\right] \quad (B6)$$

In is convenient to split the potential operator in the linear $V_1(P,t)$ and the nonlinear $V_2(\overline{P}(t),P,t)$ parts as follows

$$V_1(P,t) = -\frac{1}{4\Theta}\left[\frac{\partial U_1(P,t)}{\partial P}\right]^2 + \frac{1}{2}\frac{\partial^2 U_1(P,t)}{\partial P^2} + \frac{1}{2\Theta}\left[\frac{\partial U_1(P,t)}{\partial t}\right] \qquad (B7)$$

and

$$V_2(\overline{P}(t),P,t) =$$
$$\frac{1}{2\Theta}\frac{\partial U_2(P,t)}{\partial t} - \frac{1}{2\Theta}\frac{\partial U_1(P,t)}{\partial P}\frac{\partial U_2(P,t)}{\partial P} - \frac{1}{4\Theta}\left(\frac{\partial U_2(P,t)}{\partial P}\right)^2 + \frac{1}{2}\frac{\partial^2 U_2(P,t)}{\partial P^2} \qquad (B8)$$

resulting in Eq.(20).

## Appendix C

The error control and accuracy tests are preceded for temporal deviation of the stationary states, normalization of the probability distribution, polarization rate, and the norm conservation. Temporal behavior of the solution for stationary states $\overline{P}(0) > 0$ and $\overline{P}(0) < 0$ is estimated by deviation $(\overline{P}(t) - \overline{P}(0))/\overline{P}(0)$ and shown in Fig.B1. The time range is equivalent to a single period of $\Omega = 10^{-3}$ dimensionless driving frequency and the deviation is within one percent that is fairly good for this rough of analysis. The temporal behavior exhibits a sharp alteration (in linear time scale) and subsequent stabilization as a advancement of this approach. This result is expected as a consequence of global stability found in [8]. On other hand the result Fig.B1 exhibits a specific feature of symplectic integration, namely, reduced accuracy combined with long time stability.

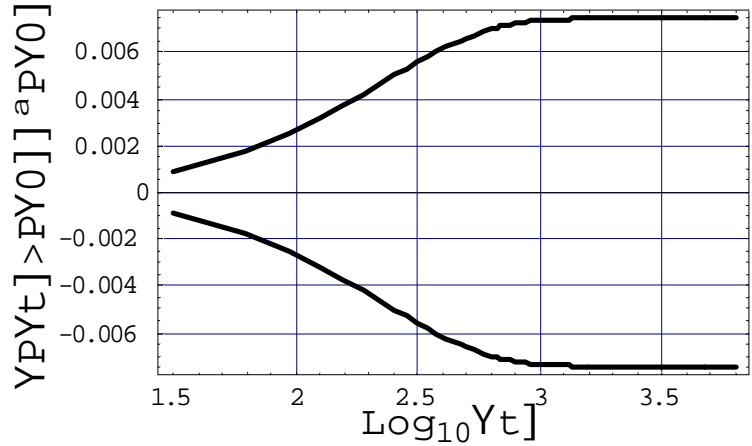

Fig.B1. Temporal behavior of the deviation $(\overline{P}(t) - \overline{P}(0))/\overline{P}(0)$ for positive $\overline{P}(0) > 0$ and negative $\overline{P}(0) < 0$ stationary states exposed by upper and bottom plot, correspondingly. Here $|\overline{P}(0)| = 0.6423$ and the increments are $\Delta P = 6/50$, $\Delta t = 1/100$.

The correction factor $\left(\int \rho(P,q,t)dP - 1\right) \ll 1$ keeping track on normalization of the probability distribution $\int \rho(P,q,t)dP = 1$ is shown in Fig.7. Its temporal behavior is similarly to this for relative deviation and exhibits both a short alteration and subsequent stabilization as expected.

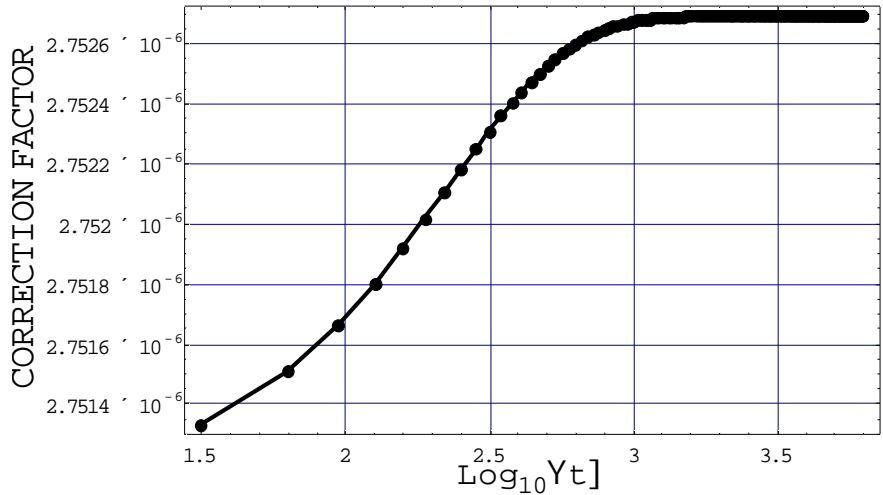

Fig.B2  Time propagation of the correction factor $\int \rho(P,q,t)dP - 1$. $\overline{P}(0) > 0$ (points), $\overline{P}(0) < 0$ (line).

The polarization rate $q_i$ shown in Fig. B3 is an essential specification of the numerical integration and comes in action from the expansion $\overline{P}_i(t_0 + \Delta t) = \overline{P}_i(t_0) + q_i \Delta t$ assumed valid for $i-th$ recursion. As expected, the polarization rate is a small quantity (at parameters of the problem), approves validity of this expansion, and reduces to zero as soon as the integration establishes.

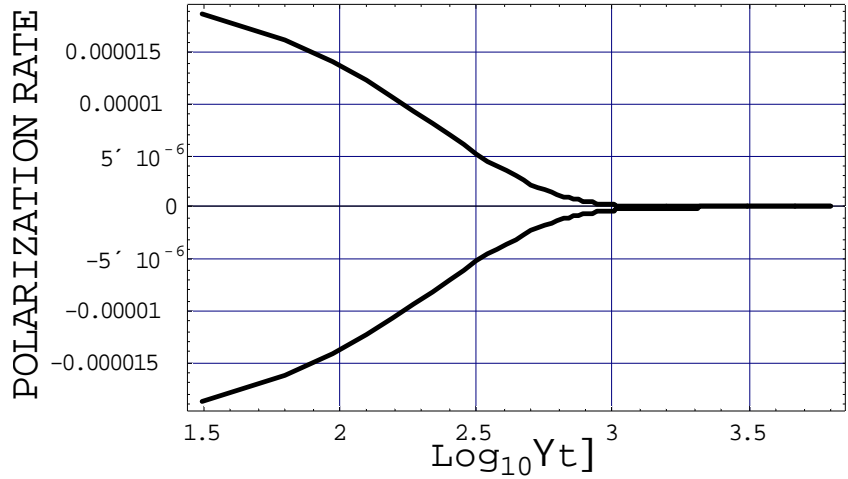

Fig. B3 Time propagation of the polarization rate (in arbitrary units).

Norm conservation being crucial for long term stability of calculations after Eq.(13) is specified by the difference of $G-$ vector norms $\|G(P,t_i)\|-\|G(P,t_i+\Delta t)\|$ within any $i-th$ recursion and is shown in Fig.B4. In comparison with the 0.3 a.u. value this difference is rather small and vanishes to zero after small nonmonotonous alterations so approving accuracy of the correctable algorithm Eq.(13) (within this set of parameters).

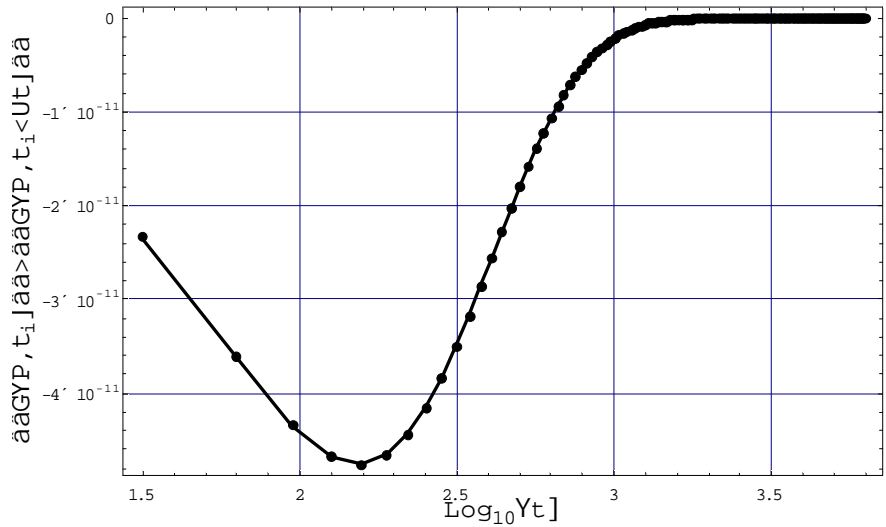

Fig. B4 Norm conservation specified by the difference $\|G(P,t_i)\|-\|G(P,t_i+\Delta t)\|$ within $i-th$ time step.

Appendix D

Decomposition of multivariate probability distribution starts with energy density for finite but great number of $N$ blocks

$$\Psi(P_1,...P_N) = \sum_i^N \left\{ \Phi_i + \frac{\varepsilon}{2}\left( (\overline{P}_{i+1}(t) - P_i)^2 + (\overline{P}_{i-1}(t) - P_i)^2 \right) \right\} \tag{D1}$$

Here the Ginzburg-Landau part is local but nonconservative

$$\Phi_i = -\frac{1}{2}P_i^2 + \frac{1}{4}P_i^4 - \lambda(t)P_i \tag{D2}$$

Kinetic equation for $i-th$ block is given by

$$\frac{\partial P_i}{\partial t} = -\frac{\partial}{\partial P_i}\left[ \Phi_i + \frac{\varepsilon}{2}\left( (\overline{P}_{i+1}(t) - P_i)^2 + (\overline{P}_{i-1}(t) - P_i)^2 \right) \right] = -\frac{\partial \Phi_i}{\partial P_i} + \varepsilon(\overline{P}_{i+1}(t) - 2P_i + \overline{P}_{i-1}(t)) \tag{D3}$$

Fokker-Planck equation for multivariate probability density is a sum over $N$ blocks

$$\dot{\rho}(\{P_1,...P_N\},t) =$$
$$-\sum_{i=1}^N \frac{\partial}{\partial P_i}\left[ -\frac{\partial \Phi_i}{\partial P_i}\rho(\{P_1,...P_N\},t) + \varepsilon(\overline{P}_{i+1}(t) - 2P_i + \overline{P}_{i-1}(t))\rho(\{P_1,...P_N\},t) \right] \tag{D4}$$
$$+ \sum_{i=1}^N \Theta_i \frac{\partial^2}{\partial P_i^2}\rho(\{P_1,...P_N\},t)$$

Stationary probability density $\dot{\rho}(\{P_1,...P_N\},t) = 0$ is given by a set of $i = [1,N]$ equations

$$\rho(\{P_1,...P_N\})\left( 2\varepsilon + \frac{\partial \Phi_i}{\partial P_i^2} \right) + 2\varepsilon P_i - \varepsilon\left( \overline{P}_{i-1} + \overline{P}_{i+1} + \frac{\partial \Phi_i}{\partial P_i} \right)\frac{\partial \rho(\{P_1,...P_N\})}{\partial P_i}$$
$$+ \Theta_i \frac{\partial^2 \rho(\{P_1,...P_N\})}{\partial P_i^2} = 0 \tag{D5}$$

That yields $i = [1,N]$ relations for the probability density

$$\rho(\{P_1,...P_N\}) \propto \exp\frac{-\Phi(P_i)}{\Theta_i}\left( 2\varepsilon + \frac{\partial \Phi_i}{\partial P_i^2} \right) + 2\varepsilon P_i - \varepsilon\left( \overline{P}_{i-1} + \overline{P}_{i+1} + \frac{\partial \Phi_i}{\partial P_i} \right)\frac{\partial \rho(\{P_1,...P_N\})}{\partial P_i}$$
$$+ \Theta_i \frac{\partial^2 \rho(\{P_1,...P_N\})}{\partial P_i^2} = 0$$